# Multiplexed control of spin quantum memories in a photonic circuit

D. Andrew Golter,[1,*] Genevieve Clark,[1,2,†] Tareq El Dandachi,[2] Stefan Krastanov,[2] Andrew J. Leenheer,[3] Noel H. Wan,[2] Hamza Raniwala,[2] Matthew Zimmermann,[1] Mark Dong,[1,2] Kevin C. Chen,[2] Linsen Li,[2] Matt Eichenfield,[3,4,‡] Gerald Gilbert,[5,ǂ] Dirk Englund[2,§]

[1]*The MITRE Corporation, 202 Burlington Road, Bedford, Massachusetts 01730, USA*
[2]*Research Laboratory of Electronics, Massachusetts Institute of Technology, Cambridge, Massachusetts 02139, USA*
[3]*Sandia National Laboratories, P.O. Box 5800 Albuquerque, New Mexico 87185, USA*
[4]*College of Optical Sciences, University of Arizona, Tucson, Arizona 85719, USA*
[5]*The MITRE Corporation, 200 Forrestal Road, Princeton, New Jersey 08540, USA*
*,†These authors contributed equally to this work.
*dagolter@mitre.org, †gclark@mitre.org, ‡meichen@sandia.gov, ǂggilbert@mitre.org, §englund@mit.edu

**Abstract**
A central goal in many quantum information processing applications is a network of quantum memories that can be entangled with each other while being individually controlled and measured with high fidelity. This goal has motivated the development of programmable photonic integrated circuits (PICs) with integrated spin quantum memories using diamond color center spin-photon interfaces. However, this approach introduces a challenge in the microwave control of individual spins within closely packed registers. Here, we present a quantum-memory-integrated photonics platform capable of (i) the integration of multiple diamond color center spins into a cryogenically compatible, high-speed programmable PIC platform; (ii) selective manipulation of individual spin qubits addressed via tunable magnetic field gradients; and (iii) simultaneous control of multiple qubits using numerically optimized microwave pulse shaping. The combination of localized optical control, enabled by the PIC platform, together with selective spin manipulation opens the path to scalable quantum networks on intra-chip and inter-chip platforms.

A network of entangled quantum memories forms an essential component of many emerging quantum technologies. These memories must be individually controllable with high fidelity, have long coherence times, and be implemented in a scalable way. Recently proposed blueprints for general-purpose quantum information processing call for solid-state spin-based quantum memories connected by switchable photonic links [1–3]. These blueprints have motivated the development of cryogenic PIC platforms [4,5] and hybrid integration with diamond color centers [6,7].

As a first step toward integrating spin memories [8–11] into such scalable PIC-based architectures, high-performance single-channel devices were demonstrated with single color centers coupled to photonic waveguides [12–17]. Recently multi-channel diamond waveguide arrays supporting multiple stable color centers were successfully incorporated into a photonic integrated circuit using a high-yield, scalable process [6,16]. To realize the potential of these color center-based devices for large-scale quantum information processing, this scalable photonic integration needs to be accompanied by scalable electron spin control. Microwave control of color center spins enables state manipulation [18,19], entanglement generation [20–23], and state transfer with long-lived nuclear memories [24–28].

Here we demonstrate microwave control over many spins integrated into a scalable piezo-optomechanical photonics platform and explore methods of scaling this control.

First we describe on-chip spin control of diamond nitrogen-vacancy (NV) centers integrated into our PICs, fabricated in a wafer-scale, 200-mm CMOS-foundry process [4,5,29,30]. Our device structure, shown in Fig. 1(a), couples NV fluorescence into on-chip silicon nitride (SiN) waveguides and edge-coupled optical fiber, enabling scalable and efficient optically networked state readout. A microwave transmission line beneath the color centers delivers a time-varying current, with a corresponding magnetic field, that allows for spin state control.

Scalability requires that individual, high-fidelity state control must be applied simultaneously to many optically networked spin qubits. However, maintaining separate microwave control channels for each qubit would dramatically increase hardware requirements. Time-frequency multiplexed microwave control offers an alternative in which many spin qubits can be controlled through a single microwave channel. In a multiplexed scheme, scalability requires selectivity in control [31–34]: i.e., it should be possible to apply a single-qubit gate on one spin while keeping crosstalk with other spins low. When applying a single-qubit rotation by operator $U_i$ on a selected spin $i$ the error on another spin $j$ is given by:

$$\epsilon_j = 1 - |\langle j_{0,1}|U_j|j_{0,1}\rangle|^2 \qquad 1)$$

where $U_j$ is the corresponding time propagator for spin $j$, and the notation indicates that the expectation value is calculated in either the ground or the excited state.

In the rotating frame of a microwave control field with angular frequency $\omega_{mw}$, Rabi frequency $\Omega(t)$, and phase $\varphi(t)$, the Hamiltonian describing the interaction with spin $i$ (which we model as a two-level system) is:

$$H_i = \frac{1}{2}\hbar[\Delta_i \sigma_z + \Omega_i(t)(\cos(\varphi_i(t))\sigma_x + \sin(\varphi_i(t))\sigma_y)] \qquad 2)$$

where $\Omega_i$ and $\Delta_i$ must be real. $\Delta_i = \omega_{+,i} - \omega_{mw}$ is the detuning from the spin transition frequency $\omega_{+,i}$ which in a static magnetic field $\vec{B}(\vec{r}_i)$, with small off-axis components, is approximately:

$$\omega_{+,i} = D_{zfs} + \gamma_{NV} B_{z,i} \qquad 3)$$

where $D_{zfs}$ is the zero-field splitting, $B_{z,i} = \hat{z} \cdot \vec{B}(\vec{r}_i)$ is the magnetic field component along the dipole axis (see Fig. 1(b)), and $\gamma_{NV}$ is the NV gyromagnetic ratio.

Applying a $\sigma_x$ rotation on spin $i$ involves setting $\Delta_i = 0$, $\Omega_i(t) = \Omega_i$, and $\varphi_i(t) = 0$ for a time $T \ll T_2$, where $T\Omega_i = \pi$ and $T_2$ is the spin coherence time. In the process, spin $j$ undergoes evolution from state $|j_0\rangle \to |j_T\rangle = U_j|j_0\rangle$. If $\Omega_j/\Delta_j \ll 1$, the state error is bounded from above by $\epsilon_j = (\Omega_j/\Delta_j)^2$ (see Supporting Information 1). This means that low crosstalk is possible if the spin transitions are well separated compared to the Rabi frequency $\Omega_j$. To achieve the $\Omega_j/\Delta_j \ll 1$ requirement for "selective frequency addressing", we use a magnetic field gradient, generated on-chip and localized to the quantum memory register, to produce a position-dependent shift in $\omega_{+,i}$. Using this method, we demonstrate frequency resolved control over spins at several different locations within a single waveguide.

On the other hand, if the spectral spacing $\Delta_j$ is too small or we want to drive the dynamics quickly, yielding a large $\Omega_j$, selective spin control is also possible by temporal modulation of $\Omega(t)$ and $\varphi(t)$. Under this scheme all spins will have non-negligible evolution under the application of the control field, but they will follow different trajectories. This opens up the possibility that, through the careful shaping of $\Omega(t)$ and $\varphi(t)$ via the tools of optimal control theory, the spins can be guided to a desired end state, effectively executing selective control even when the conditions for full frequency resolution are not met. We conclude by considering, theoretically and experimentally, this simultaneous control using optimally shaped microwave pulses.

We performed experiments on naturally occurring NV⁻ centers in an 8-waveguide single crystal diamond "quantum micro-chiplet" (QMC) integrated over a microwave transmission line on a silicon-nitride (SiN) PIC, as seen in Fig. 1(a) and (c). The diamond QMC waveguides (see Ref. [6] for fabrication) and SiN waveguides on the PIC (see Ref. [4]) couple optically via inverse tapering, as detailed previously [16]. Details of the waveguide dimensions and coupling efficiency are described in Supporting Information 2. Optical excitation of the color centers occurs through free-space, perpendicular to the chip, while fluorescence is collected from the SiN waveguides into an edge-coupled optical fiber. With the excitation laser positioned at location 1 in Fig. 1(a), we observe a clear NV zero-phonon-line when collecting fluorescence through the PIC as shown in Fig 1(d).

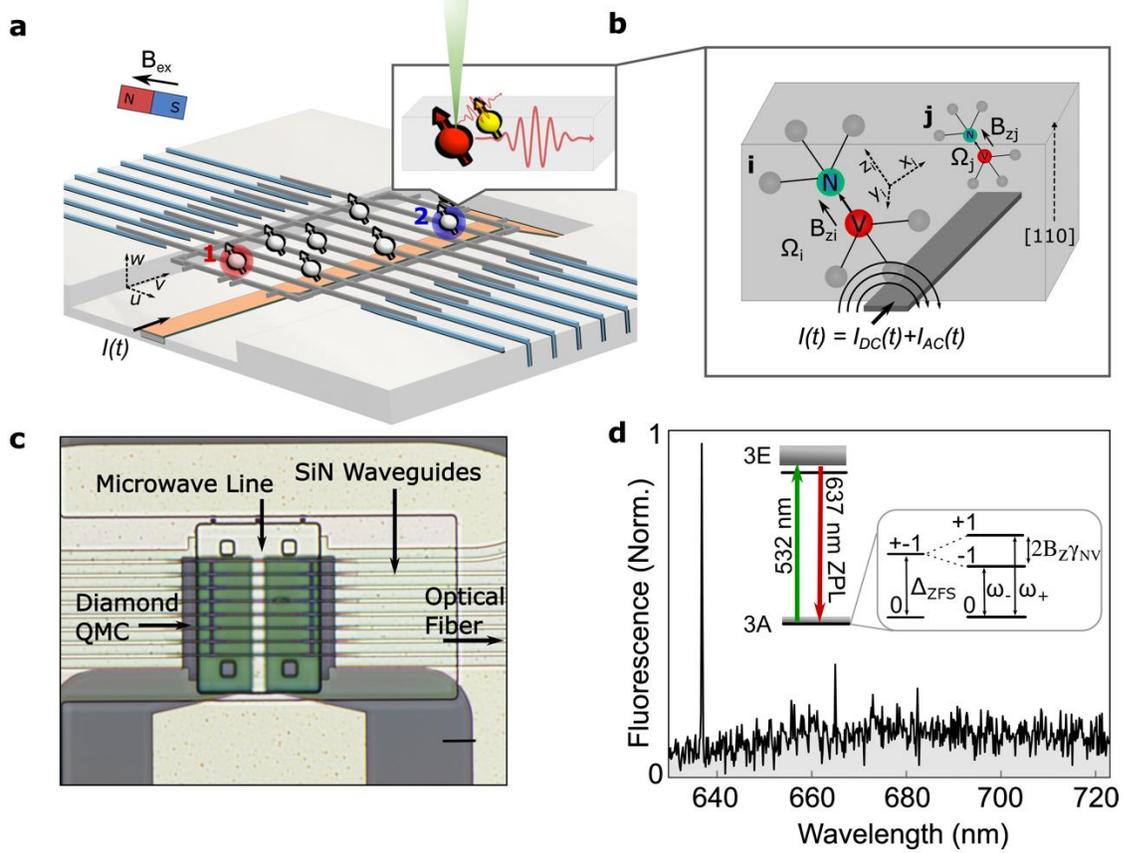

**FIG. 1. (a)** Device structure showing spins in a QMC heterogeneously integrated into a PIC. A free-space laser excites NV centers in the diamond waveguides while NV fluorescence couples to the diamond waveguide mode. **(b)** A current $I(t)$ supplied through the microwave line is used to manipulate NV spin qubits in the QMC. **(c)** Optical microscope image of the PIC showing integrated wire for microwave delivery below the diamond QMC. **(d)** Low-temperature photoluminescence (PL) spectrum of NV centers in the QMC. Inset shows the energy level diagram for NV centers subject to a magnetic field $B_z$.

Figure 1(b) illustrates our approach to spin control of NV centers in the QMC using the microwave line. A current with both an AC and a DC component, $I(t) = I_{DC} + I_{AC}(t) \sin(\omega_{mw} t + \phi_{mw})$ where $I_{AC}(t)$ varies slowly compared with $\omega_{mw}$, is applied through the microwave line creating a spatially dependent DC magnetic field, $\vec{B}_{DC}(\vec{r})$, and a time and spatially dependent AC magnetic field, $\vec{B}_{AC}(t,\vec{r}) \sin(\omega_{mw} t + \phi_{mw})$, with $|\vec{B}_{DC}| \propto I_{DC}$ and $|\vec{B}_{AC}| \propto I_{AC}$. In addition, an external magnet provides a constant magnetic field, $\vec{B}_{ext}$, that is spatially invariant at the length-scale of the QMC. The spin $|\pm 1\rangle$ sublevels for NV $i$ at position $\vec{r}_i$ undergo a Zeeman shift (see Eq. 3) yielding spin transitions at

$$\omega_{\pm,i} = D_{zfs} \pm \gamma_{NV}[B_{ext,z} + B_{DC,z}(\vec{r}_i)] \qquad 4)$$

as shown in Fig. 1(d) inset. When $\omega_{mw} = \omega_{\pm,i}$, the spin transition is driven at the Rabi frequency, $\Omega_i(t) = \frac{1}{\sqrt{2}}\gamma_{NV}B_{AC,xy}(t,\vec{r}_i)$, where $B_{AC,xy}$ is the component of the AC magnetic field perpendicular to the dipole axis.

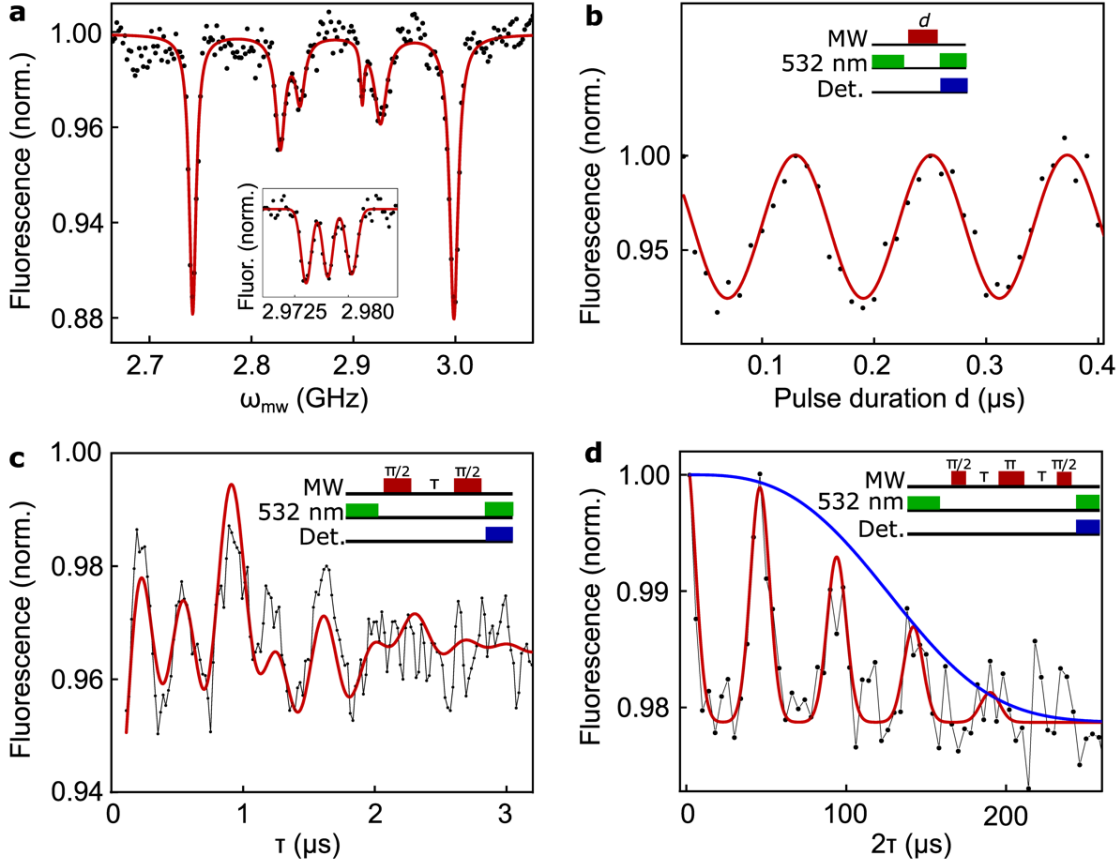

**FIG. 2. (a)** ODMR spectrum measured in a static uniform DC magnetic field under continuous microwave excitation. Inset: ODMR spectrum measured under low excitation power reveals triplet hyperfine structure due to coupling to the nitrogen nuclear spin. **(b)** Spin-dependent fluorescence as a function of microwave pulse duration shows Rabi oscillations when the microwave field is resonant with a particular spin transition. **(c)** Ramsey fringes reveal a $T_2^*$ of 1.7 µs. The red line is a fit using three exponentially decaying sine waves with frequencies given by the detuning for each of the hyperfine states. **(d)** Hahn echo sequence reveals a $T_2$ of 150 µs. The echo signal collapses and revives at the frequency of the Larmor precession of nearby $^{13}C$ nuclear spins.

We begin by setting $I_{DC} = 0$. Figure 2(a) shows optically detected magnetic resonance (ODMR) when optically pumping location 2 indicated in Fig. 1(a). We observe three resonance pairs, indicating at least three NV centers at this location in the QMC, with different orientations relative to $\vec{B}_{ext}$. The inset shows the spin-triplet hyperfine structure of the prevalent $^{14}NV$ isotope. With the microwave field tuned to one of the resonances, $\omega_{mw}/2\pi = 3.00$ GHz, measurements following the pulse sequence shown in the inset of Fig. 2(b) produce Rabi oscillations at $\Omega_i/2\pi = 7.5$ MHz. We next apply this spin control to Ramsey fringe (Fig. 2(c)) and Hahn echo (Fig. 2(d)) measurements, yielding coherence times for this spin memory of $T_2^* = 1.7$ µs and $T_2 = 150 \pm 5$ µs, typical for NV spins in diamond nanostructures [16]. For the remainder of the measurements in this paper, we consider only the NV subpopulation corresponding to the feature at 3.00 GHz in Fig. 2(a), and set $\hat{z}$ along the corresponding dipole axis.

To demonstrate selective frequency addressing, we now set $I_{DC} \neq 0$, giving $\omega_{\pm,i}$ a spatial dependence (Eq. 4). Assuming all spins are initialized to $|0\rangle$, the crosstalk error on spin $j$ under a π-pulse on spin $i$ is $\epsilon_j = 1 - \langle 0|U_j|0\rangle^2$. Figure 3(a) plots simulated $\epsilon_j$ with $I_{DC} = 0$ mA and 150 mA, for a target spin $i$ located at $u = 1.5$ μm and a π-pulse with $\Omega_i/2\pi = 10$ MHz to match experiment. We determine the spatial dependence of $\vec{B}_{DC}$ and $\vec{B}_{AC}$ from finite-element simulations based on the current in the microwave line. We then calculate $B_{DC,z}$ and $B_{AC,xy}$ for an NV dipole axis with $\theta_w = 54.7°$ for NVs in [100] oriented diamond and with $\theta_u = 41.0°$ set to be consistent with our experimental data for the subpopulation of NV centers considered in this work. For $I_{DC} = 150$ mA, $\epsilon_j$ is expected to drop significantly within a few micrometers of the target spin, while for $I_{DC} = 0$ mA, $\epsilon_j$ remains large at locations many micrometers from the target qubit (Fig. 3(b)i). The spatially varying Zeeman shift, Fig. 3(b)ii, has led to a position-dependent $\omega_j$, yielding a $\Delta_j$ sufficiently large to satisfy the condition $\Omega_j/\Delta_j \ll 1$ even for closely spaced spins

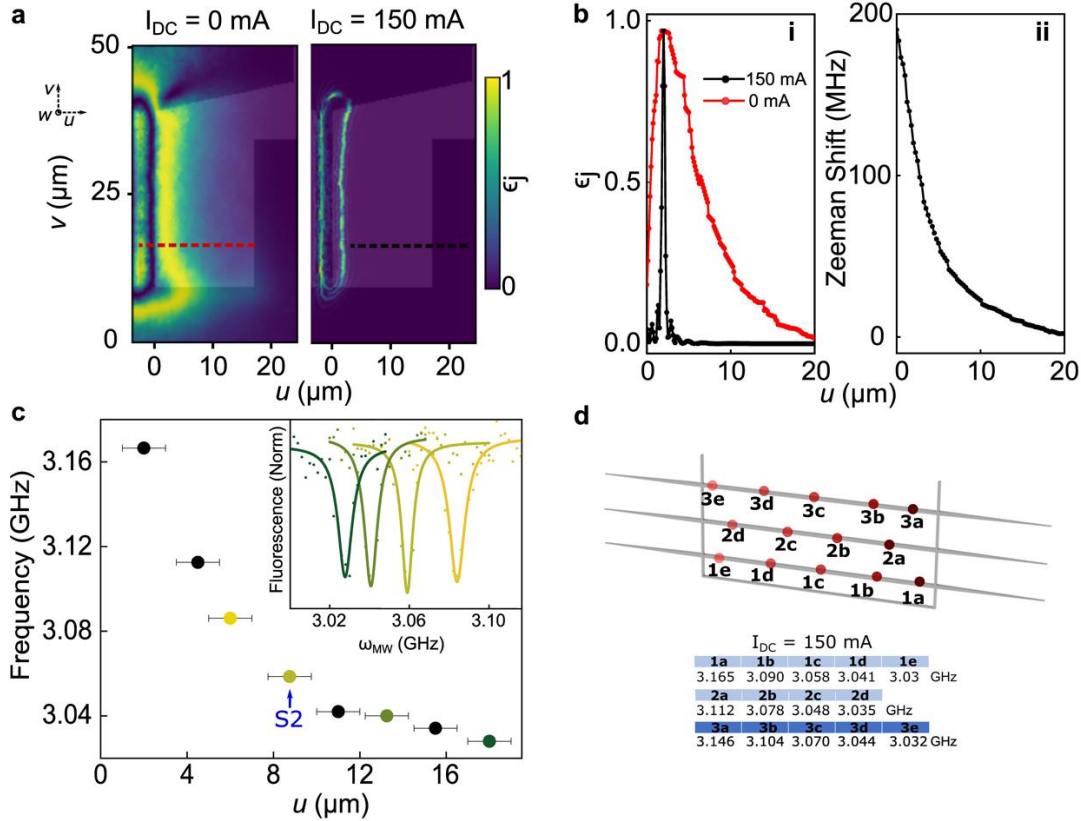

**FIG. 3. (a)** Simulated state error $\epsilon_j$ with $I_{DC}$ of 0 and 150 mA. **(b) i)** Line cuts taken from panel a, showing simulated $\epsilon_j$ as a function of $u$. **ii)** Calculated Zeeman shift as a function of $u$ for $I_{DC} = 150\,mA$. **(c)** Measured transition frequency for spins at different locations along a single diamond waveguide, with $I_{DC} = 150\,mA$, exhibiting a position dependent Zeeman shift. Error bars correspond to uncertainty in location along the QMC. Inset: A sample of ODMR measurements used to determine the transition frequencies in the main plot. The spatially varying Zeeman shift allows selective excitation of multiple spins. **(d)** Spins at different locations in the QMC are mapped to specific microwave frequency "addresses" for a given DC current. Experimentally measured frequency addresses are shown for multiple locations along three different diamond waveguides in our device.

To demonstrate this experimentally, we apply a DC current, $I_{DC} = 150$ mA, to the microwave line. In ODMR measurements, the NV orientation corresponding to the 3 GHz feature at spot 2 (measured in Fig. 2) shifts to 3.06 GHz, as seen in the labeled measurement in Fig. 3(c). Mapping the transition frequency for this orientation subpopulation along the length of a single diamond waveguide (along coordinate $u$ as shown in Fig. 1(a) and Fig. 3(a)), the Zeeman shift increases to over 160 MHz at $u = 2$ µm. This allows multiple spins in the same waveguide to be resolved, as shown in the inset. Figure 3(d) summarizes the potential of this technique to enable selective addressing of single spins within a dense register. The spatial variation of $\vec{B}_{DC}$ allows each spin in the QMC to be mapped to a specific address, either in $\omega_{mw}$ or $I_{DC}$. Figure 3(d) shows the measured $\omega_{mw}$ address for spins in three channels of the QMC at an $I_{DC}$ of 150 mA.

The NV electron spin is coupled to the nitrogen nuclear spin. For these simulations and measurements we have considered the case where manipulation of the electron spin is independent of the nuclear spin state. This leads to a tradeoff between the control fidelity (improved by larger $\Omega_i$) and the reduction in $\epsilon_j$ (improved by smaller $\Omega_j$). For a discussion of the case where there is no hyperfine coupling or the control is nuclear spin dependent, see Supporting Information 6.

The scalability of this selective control scheme is limited by the steepness of the field gradient due to the requirement that $\Delta_j \gg \Omega_j$. Scaling to hundreds of spins and beyond calls for a technique to selectively control spins with more closely spaced transition frequencies. Using optimal control techniques, we can relax the requirement that $\Delta_j \gg \Omega_j$, pushing the limits of selective control even further.

Consider the special case of a spin flip operation performed on a single target spin $i$, and a single neighbor spin $j$ that we wish to remain unchanged, with $\omega_i - \omega_j \neq 0$ and with both spins initialized into their respective ground states, $|0_i\rangle$ and $|0_j\rangle$. (See Supporting Information 7 for examples of other control scenarios.)

Our optimal control technique works by shaping the time dependent in-phase, $\mathcal{I}(t) = \frac{1}{2\pi}\Omega(t)\cos(\varphi(t))$, and out-of-phase, $\mathcal{Q}(t) = \frac{1}{2\pi}\Omega(t)\sin(\varphi(t))$ components of a microwave control pulse. This pulse is composed of $m$ discrete time steps, with $\Omega$ and $\varphi$ constant during each step. We then apply gradient descent [35–39] to minimize a cost function, $f(\mathcal{I},\mathcal{Q}) \equiv (1 - \epsilon_i) + \epsilon_j + R$, which includes the state error for each qubit, $\epsilon_i = 1 - |\langle 0_i| \prod_{l=1}^{m} U_{i,l}(\mathcal{I},\mathcal{Q}) |0_i\rangle|^2$ and $\epsilon_j = 1 - |\langle 0_j| \prod_{l=1}^{m} U_{j,l}(\mathcal{I},\mathcal{Q}) |0_j\rangle|^2$, where $U_{i,l}$ and $U_{j,l}$ are the evolution operators for a given time step $l$. Also included in the cost function is a regularization term, $R = \lambda \sum_{l=1}^{m-1}[|\mathcal{I}(t_l) - \mathcal{I}(t_{l+1})| + |\mathcal{Q}(t_l) - \mathcal{Q}(t_{l+1})|]$, weighted by $\lambda$ (real, typically set to $< 1$ MHz$^{-1}$), that is included to ensure that the pulse is smooth enough to be generated by the microwave source.

As mentioned above, the coupling between the NV electron spin and the nitrogen nuclear spin produces a hyperfine triplet with splittings of 2.2 MHz. The nuclear spin is in a mixed state, so to include this hyperfine interaction we model and average over 3 two-level systems for both spin $i$ and spin $j$. Optimally shaped pulses can be designed to flip one spin selective on the corresponding nuclear spin state. However, experimentally we consider the case where manipulation of the electron spin is independent of the nuclear spin state. While this placed a limit on the effectiveness of frequency addressing, we show that our use of optimal control pulses can mitigate this limitation as well.

We experimentally demonstrate optimized control for the case where $\Delta_j/2\pi = 1.1$ MHz. This represents an order of magnitude improvement in the minimum frequency separation needed for selective control compared to the $\Delta_j/2\pi > 20$-$30$ MHz needed for frequency resolution in Section III. Figure 4(a) shows the $\mathcal{I}(t)$ and $\mathcal{Q}(t)$ components of a control pulse designed using the optimization method described above. The calculated spin evolution during the application of this pulse is plotted in Fig. 4(b) for each of the hyperfine states of $i$ and $j$, with both spins starting at $|0\rangle$ and with an arrow indicating the end state.

We choose two spatially separated sites in the QMC and adjust $I_{DC}$ to achieve $\Delta_j/2\pi = 1.1$ MHz for those two sites. Spins at both sites are optically initialized into the $|0\rangle$ state. In Fig. 4(c) we show, for each spin, the population found to be in the $|1\rangle$ state after the application of a microwave pulse, as measured by the change in fluorescence. This is normalized to the case of a rectangular π-pulse on resonance with that spin. A short, rectangular pulse can flip spin $i$ with high fidelity, however the crosstalk with spin $j$ is significant ($\epsilon_j > 0.9$). A longer rectangular pulse can achieve better resolution, but it becomes hyperfine state selective, flipping at most a third of the population. In comparison, the optimized pulse successfully flips $i$ while significantly reducing crosstalk with $j$.

In our calculations this pulse shape achieves $\epsilon_j = 0.006$ and $\epsilon_i = 0.995$, however this is extremely sensitive to $\Delta_j$ (Supporting Information 8) as well as to pulse amplitude. The larger crosstalk in our measurement is likely due to a limit in the precision with which these parameters were set, as well as to limitations in our spin state measurements (see below).

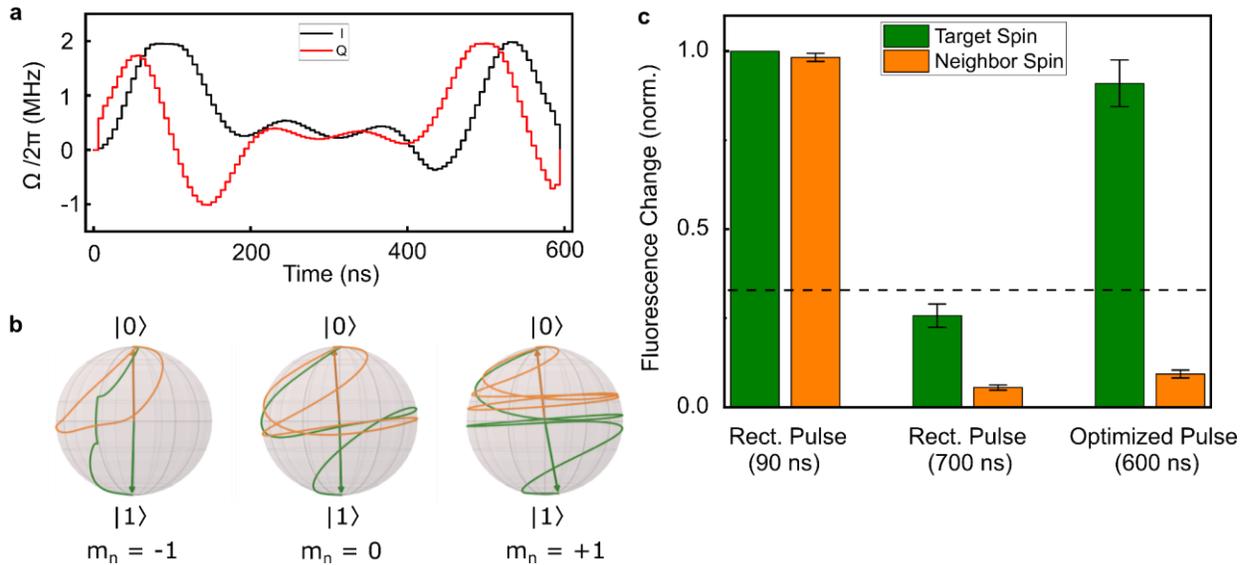

**FIG. 4.** (a) Optimized pulse shape for selectively applying a π-rotation on spin i while leaving spin j unchanged, where $\omega_i - \omega_j = 1.1$ MHz. The amplitudes of the in-phase, $I$ (black), and out-of-phase, $Q$ (red), components of the microwave signal are shown as a function of time. (b) Calculated paths traced out on the Bloch sphere by spin i (green) and spin j (orange) during application of the optimized pulse. The hyperfine coupling to the nuclear spin has been included, and the spin evolution is shown for each of the three nuclear spin states, $m_n$. (c) Comparison of the crosstalk caused by rectangular control pulses and the optimized control pulse. The change in the spin-dependent fluorescence resulting from the application of the respective pulses is shown for each spin, normalized to the case of a rectangular pulse on resonance with that spin. The dashed line indicates the maximum expected fluorescence change (1/3) if the spin is flipped for only one hyperfine case. Error bars indicate the standard error from repeated measurements.

We have demonstrated scalable methods for controlling spins in dense registers with low crosstalk, in a CMOS compatible PIC platform. Using an on-chip microwave signal line, we control the state of NV center spins hosted in optically integrated diamond waveguides. We also use this line to generate a local magnetic field gradient, enabling frequency addressing of multiple spins within a single waveguide for individual control. Finally, we use tailored microwave pulses optimized through a gradient-based search to selectively control spins where the requirement for frequency resolution is not met, demonstrating control with minimal crosstalk for frequency separations as small as 1.1 MHz.

These control methods are applicable to large numbers of spins (see Supporting Information 6), are fully coherent, and are compatible with state transfer to nuclear memories. While the measurements shown here are limited to $\pi$-rotations and state population readout, the selective spin control is effective for arbitrary rotations (see Supporting Information 7).

To aid in optical integration and to explore the spatial dependence of the Zeeman shift, we used a diamond QMC with a moderately high density of color centers throughout. This made spin state measurements challenging due to a large position dependent background from color centers of the wrong orientation and to the existence of multiple color centers of the correct orientation at different locations within the optical excitation spot. These limitations will be mitigated by using lower density samples and resonant optical addressing [6,31].

Our device used a single, straight line for generating the magnetic field gradient. In the future, more sophisticated line designs could allow for tailored gradient shapes and steepness, and multiple lines would enable tuning of the field direction for alignment with color centers.

The results shown here represent an important step toward the development of scalable quantum networks with nodes containing large numbers of spin qubits that can be selectively manipulated with high fidelity while being interfaced with on and off-chip optics.

**Acknowledgments:**

Major funding for this work is provided by The MITRE Corporation for the Quantum Moonshot Program. N.H.W. acknowledges support from the Army Research Laboratory (ARL) Center for Distributed Quantum Information (CDQI) program W911NF-15-2-0067. H.R. acknowledges support from the NDSEG Fellowship and the NSF Center for Ultracold Atoms (CUA). K.C.C. acknowledges additional funding support by the National Science Foundation RAISE-TAQS (Grant No.1839155). L.L. acknowledges funding from NSF QISE-NET Award (DMR-1747426) and the ARO MURI W911NF2110325. M.E. performed this work, in part, with funding from the Center for Integrated Nanotechnologies, an Office of Science User Facility operated for the US Department of Energy Office of Science. D.E. acknowledges the National Science Foundation (NSF) Engineering Research Center for Quantum Networks (CQN), awarded under cooperative agreement number 1941583.

**Author Contributions:**

D.A.G. and G.C., with assistance from H.R., built the experimental setups and performed the experiments. T.E.D. and S.K. performed the gradient descent numerical analysis to design the mw pulse shapes for optimal control and M.Z. assisted in experimental microwave pulse generation. G.G. provided additional theory support. N.H.W., K.C.C., and L.L. designed and fabricated the diamond micro-chiplets. G.C. and N.H.W., with assistance from M.D., A.J.L., and M.E., designed the PIC for diamond integration. M.E. developed the PIC platform and concept of operation. M.E. and A.J.L. developed the fabrication methodology of the PIC platform. D.E., D.A.G. and G.C. conceived the experiments. G.G., M.E., and D.E. supervised the project. D.A.G. and G.C. wrote the manuscript with input from all authors.